\RequirePackage[switch]{lineno_babar_revtex}
\documentclass[twocolumn,showpacs,aps,prl,superscriptaddress]{revtex4}

\catcode`\@=11\relax
\def\@makecol{%
 \setbox\@outputbox\vbox{%
  \boxmaxdepth\@maxdepth
 \protected@write\@auxout{}{%
 \string\@LN@col{\@ifnum{\pagegrid@cur=\@ne}{1}{2}}
      }%
  \@tempdima\dp\@cclv
  \unvbox\@cclv
  \vskip-\@tempdima
 }%
 \xdef\@freelist{\@freelist\@midlist}\global\let\@midlist\@empty
 \@combinefloats
 \@combineinserts\@outputbox\footins
  \set@adj@colht\dimen@
  \count@\vbadness
  \vbadness\@M
  \setbox\@outputbox\vbox to\dimen@{%
   \@texttop
   \dimen@\dp\@outputbox
   \unvbox\@outputbox
   \vskip-\dimen@
   \@textbottom
  }%
  \vbadness\count@
 \global\maxdepth\@maxdepth
}%
\def\balance@two#1#2{%
\outputdebug@sw{{\tracingall\scrollmode\showbox#1\showbox#2}}{}%
 \setbox\@ne\vbox{%
  \@ifvoid#1{}{%
   \unvcopy#1\recover@footins
   \@ifvoid#2{}{\marry@baselines}%
  }%
  \@ifvoid#2{}{%
   \unvcopy#2\recover@footins
  }%
 }%
 \dimen@\ht\@ne\divide\dimen@\tw@
 \dimen@i\dimen@
 \vbadness\@M
 \vfuzz\maxdimen
 \loopwhile{%
  \dimen@i=.5\dimen@i
  \outputdebug@sw{\saythe\dimen@\saythe\dimen@i\saythe\dimen@ii}{}%
  \setbox\z@\copy\@ne\setbox\tw@\vsplit\z@ to\dimen@
  \setbox\z@ \vbox{%
 \protected@write\@auxout{}{%
 \string\@LN@col{\@ifnum{\pagegrid@cur=\@ne}{1}{2}}
      }%
   \unvcopy\z@
   \setbox\z@\vbox{\unvbox\z@ \setbox\z@\lastbox\aftergroup\vskip\aftergroup-\expandafter}\the\dp\z@\relax
  }%
  \setbox\tw@\vbox{%
   \unvcopy\tw@
   \setbox\z@\vbox{\unvbox\tw@\setbox\z@\lastbox\aftergroup\vskip\aftergroup-\expandafter}\the\dp\z@\relax
  }%
  \dimen@ii\ht\tw@\advance\dimen@ii-\ht\z@
  \@ifdim{\dimen@i>.5\p@}{%
   \advance\dimen@\@ifdim{\dimen@ii<\z@}{}{-}\dimen@i
   \true@sw
  }{%
   \@ifdim{\dimen@ii<\z@}{%
    \advance\dimen@\tw@\dimen@i
    \true@sw
   }{%
    \false@sw
   }%
  }%
 }%
 \outputdebug@sw{\saythe\dimen@\saythe\dimen@i\saythe\dimen@ii}{}%
\@ifdim{\ht\z@=\z@}{%
\@ifdim{\ht\tw@=\z@}{%
\true@sw
}{%
\false@sw
}%
}{%
\true@sw
}%
{%
}{%
\ltxgrid@info{Unsatifactorily balanced columns: giving up}%
\setbox\tw@\box#1%
\setbox\z@ \box#2%
}%
 \setbox\tw@\vbox{\unvbox\tw@\vskip\z@skip}%
 \setbox\z@ \vbox{\unvbox\z@ \vskip\z@skip}%
 \set@colroom
\dimen@\ht\z@\@ifdim{\dimen@<\ht\tw@}{\dimen@\ht\tw@}{}%
\@ifdim{\dimen@>\@colroom}{\dimen@\@colroom}{}%
 \outputdebug@sw{\saythe{\ht\z@}\saythe{\ht\tw@}\saythe\@colroom\saythe\dimen@}{}%
\setbox#1\vbox to\dimen@{\unvbox\tw@\unskip\raggedcolumn@skip}%
\setbox#2\vbox to\dimen@{\unvbox\z@ \unskip\raggedcolumn@skip}%
\outputdebug@sw{{\tracingall\scrollmode\showbox#1\showbox#2}}{}%
}%
\catcode`\@=12\relax

\usepackage{palatino}
\usepackage{graphicx}
\usepackage{dcolumn}
\usepackage{amsmath}
\usepackage{amsfonts}
\usepackage{amssymb}
\usepackage{amsbsy}
\usepackage{epsfig}
\usepackage{verbatim}

\RequirePackage{xspace}

\input pubboard/babarsym

\providecommand{\theLumi}  {\ensuremath{347.3}\xspace}

\providecommand{\thislimitEpsBayes}{\ensuremath{0.68\%}\xspace}
\providecommand{\thislimitRatBayes}{\ensuremath{0.69\%}\xspace}
\providecommand{\thislimitMoreEpsBayes}{\ensuremath{0.84\%}\xspace}

\providecommand{\datEps}{\ensuremath{(-0.55\pm0.58\text{(stat)}^{+0.08}_{-0.55}\text{(syst)})\%}\xspace}

\providecommand{\Km}{\ensuremath{K^{-}}\xspace}
\providecommand{\Kpm}{\ensuremath{K^{\pm}}\xspace}

\providecommand{\piz}{\ensuremath{\pi^0}\xspace}
\providecommand{\pip}{\ensuremath{\pi^+}\xspace}
\providecommand{\pim}{\ensuremath{\pi^-}\xspace}

\providecommand{\Br}{\ensuremath{\mathcal{B}}\xspace}
\providecommand{\fbarn}    {\ensuremath{\text{fb}^{-1}}\xspace}

\providecommand{\epem}  {\ensuremath{e^+e^-}\xspace}

\providecommand{\om}   {\ensuremath{\omega}\xspace}

\providecommand{\ta}   {\ensuremath{\tau}\xspace}

\providecommand{\tm}   {\ensuremath{\tau^-}\xspace}

\providecommand{\nut}  {\ensuremath{\nu_{\tau}}\xspace}
\providecommand{\qqbar}{\ensuremath{q\bar{q}}\xspace}

\providecommand{\tauom}  {\ensuremath{\tm\to\omega\pim\nut}\xspace}

\providecommand{\taub}  {\ensuremath{\tm\to b_{1}^-\nut}\xspace}
\providecommand{\bom}   {\ensuremath{b_1^-\to\omega\pim}\xspace}

\providecommand{\ompppz}{\ensuremath{\om\to\pip\pim\piz}\xspace}
\providecommand{\pppz}{\ensuremath{\pip\pim\piz}\xspace}
\providecommand{\ppppz}{\ensuremath{\pim\pim\pip\piz}\xspace}
\providecommand{\tppppz}{\ensuremath{\tm\to\pim\pim\pip\piz\nut}\xspace}
\providecommand{\mtpz}{\ensuremath{m(\ppppz)}\xspace}
\providecommand{\mopz}{\ensuremath{m(\pppz)}\xspace}
\providecommand{\epsil}{\ensuremath{\epsilon}\xspace}

\providecommand{\thet}{\ensuremath{\theta_{\om\pi}}\xspace}
\providecommand{\costhet}{\ensuremath{\cos\thet}\xspace}

\providecommand{\MeVcc}{\ensuremath{\mbox{MeV/}c^2}\xspace}

\providecommand{\omppz}{\ensuremath{\om\pim\piz}\xspace}
\providecommand{\tauomppz}{\ensuremath{\tm\to\om\pim\piz\nut}\xspace}

\newcommand{\BABARPubYear}    {09}
\newcommand{\BABARPubNumber}  {005}

\newcommand{\SLACPubNumber} {13579}

\long\def\inst#1{\par\nobreak\kern 4pt\nobreak
    {\it #1}\par\vskip 10pt plus 3pt minus 3pt}

\begin{document}


\noindent
\babar-PUB-\BABARPubYear/\BABARPubNumber \\
SLAC-PUB-\SLACPubNumber \\


\title{
  \boldmath Search for Second-Class Currents in {\boldmath$\tauom$}
}

%
\author{B.~Aubert}
\author{Y.~Karyotakis}
\author{J.~P.~Lees}
\author{V.~Poireau}
\author{E.~Prencipe}
\author{X.~Prudent}
\author{V.~Tisserand}
\affiliation{Laboratoire d'Annecy-le-Vieux de Physique des Particules (LAPP), Universit\'e de Savoie, CNRS/IN2P3,  F-74941 Annecy-Le-Vieux, France}
\author{J.~Garra~Tico}
\author{E.~Grauges}
\affiliation{Universitat de Barcelona, Facultat de Fisica, Departament ECM, E-08028 Barcelona, Spain }
\author{M.~Martinelli$^{ab}$}
\author{A.~Palano$^{ab}$ }
\author{M.~Pappagallo$^{ab}$ }
\affiliation{INFN Sezione di Bari$^{a}$; Dipartimento di Fisica, Universit\`a di Bari$^{b}$, I-70126 Bari, Italy }
% ...
\collaboration{The \babar\ Collaboration}
\noaffiliation

\date{\today}

\begin{abstract}
We report an analysis of $\tau^-$ decaying into $\omega\pi^-\nu_\tau$ with 
$\omega\to\pi^+\pi^-\pi^0$ using a data sample containing nearly 320 
million $\tau$ pairs 
collected with
the \babar\ detector at the PEP-II 
$B$-Factory. We find no evidence 
for second-class currents and we
set an upper limit 
of $0.69\%$ at 90\% confidence level
for the fraction of second-class currents in this decay mode.

\end{abstract}


\maketitle

Hadronic weak currents can be classified as either 
first- or second-class
depending on the spin $J$, parity $P$ and G-parity $G$ of the 
final hadronic system \cite{scc}.
In the 
Standard Model, first-class currents (FCC) in $\tau$ decays
have $J^{PG} = 0^{++}, 0^{--}, 1^{+-}$ or $1^{-+}$, and are expected to
dominate. 
Second-class currents (SCC) have $J^{PG} = 0^{+-}, 0^{-+}, 1^{++}$ 
or $1^{--}$, and 
are associated with a decay constant proportional to the mass
difference between up and down quarks. 
Thus they are  expected to vanish in the limit of
perfect isospin symmetry. 
SCC searches have taken place extensively in nuclear $\beta$ decay 
experiments~\cite{nuclear1, nuclear2}, with no confirmed observations. 
This letter presents a search for SCC in \tauom\ decays with \ompppz,
based on studying the angular distributions of final-state particles.

The decay $\tm\to\omega\pim\nut$\cite{charge}
is expected to proceed predominantly through FCC
mediated by the $\rho$ resonance.
This decay may also potentially proceed through SCC with 
$J^{PG}=0^{-+}$ or $1^{++}$. The latter may be 
mediated by $b_1$(1235)~\cite{leroy} with $\taub\to\om\pim\nut$.
The decay \bom occurs through S- and D-waves~\cite{pdg}, as compared
to a P-wave for FCC. 
Different alignments of \om spin result
in different angular distributions of the final state particles.
The expected distributions of \costhet, $F(\costhet)$, for all possible 
spin-parity
states of the final state particles are listed in Table~\ref{table:sccshape}, 
where \thet is the angle between the normal to the \om decay
plane and the direction of the remaining $\pi$ in the \om rest frame.
The existing measurements of the
angular distribution in \tauom are
consistent with having only the P-wave contribution, and the present 
upper limit is 5.4\% for the ratio of SCC to FCC contributions
at 90\% confidence level (CL)~\cite{alephscc,cleoscc}. 

\begin{table}[h!]
\caption{Expected angular distributions, $F(\costhet)$, for possible
spin-parity states in the decay \tauom. $L$ is the orbital angular
momentum.}
\begin{center}
\begin{tabular}{c|c|c}
\hline
$J^{P}$ & $L$  & $F(\costhet)$ \\
\hline
$1^-$ & 1   & $(1-\mbox{cos}^2\thet)$ \\
\hline
$0^-$ & 1   & $\mbox{cos}^2\thet$ \\
$1^+$ & 0   & $1$\\
$1^+$ & 2   & $(1+3\mbox{ cos}^2\thet)$\\
\hline
\end{tabular}
\end{center}
\label{table:sccshape}
\end{table}

This analysis is based on data recorded 
by the \babar\ detector \cite{babar} at the \pep2\ asymmetric-energy 
\epem\ storage rings operated at the SLAC National Accelerator Laboratory. 
The data sample consists of \theLumi\ \fbarn recorded at
a center-of-mass energy of $10.58\gev$.
With a cross section for \ta\ pairs
of $\sigma_{\tau\tau} = (0.919\pm0.003)$ nb \cite{tauxsect,kk2f},
this data sample contains nearly 320 million pairs of $\tau$ decays.
 
The \babar\ detector is described in detail in Ref.~\cite{babar}.
Charged-particle tracks are measured with a 5-layer
double-sided silicon vertex tracker (SVT) together with a 40-layer drift 
chamber (DCH)
inside a 1.5-T superconducting solenoid magnet.
An electromagnetic calorimeter (EMC) consisting of 6580 CsI(Tl) 
crystals is used for identification of electrons and photons.
Charged hadrons are identified by a ring-imaging Cherenkov detector
in combination with energy-loss measurements  
($dE/dx$) in the SVT and the DCH.
An instrumented magnetic-flux return (IFR) provides muon 
identification.

Monte Carlo simulation is used to estimate 
the signal efficiencies and background contamination.
{\tt KK2f} \cite{kk2f} is used to generate \ta pairs with
the decays of the \ta\ leptons  modeled by {\tt Tauola} \cite{tauola}.
Continuum \qqbar\ events are simulated using {\tt JETSET} \cite{jetset}.
Final-state radiative effects are generated for all 
decays using {\tt Photos} \cite{photos}.
The detector response is simulated 
with {\tt GEANT4} \cite{geant}, and the Monte Carlo events are
reconstructed in the same manner as data. 

Since \ta\ pairs are produced back-to-back in 
the \epem\ center-of-mass frame, each event is divided into two
hemispheres according to the thrust axis \cite{thrust}, calculated 
using all reconstructed charged particles. Candidate events in this 
analysis are required to 
have a ``1-3 topology'', where one track is in one hemisphere 
(tag hemisphere) and three tracks are in the other hemisphere 
(signal hemisphere). Events with four well-reconstructed tracks 
and zero 
net charge 
are selected for further analysis.
The polar angles of all four tracks 
and the neutral clusters used in \piz reconstruction
are required to be within the calorimeter acceptance 
range. Events are rejected if the invariant mass of any
pair of oppositely charged tracks, assuming electron mass hypotheses,
is less than 90 \MeVcc, as these tracks are likely to be 
from photon conversions in the detector material. 

The charged particle found in the tag hemisphere 
must be either an electron or a muon 
candidate. Electrons are identified using the ratio of 
calorimeter energy to track momentum ($E/p$), the shape of 
the shower in the calorimeter, and $dE/dx$.
Muons are identified by signals in the IFR and small energy 
deposits in the calorimeter consistent with expectation
for a minimum-ionizing particle. 
Charged particles found in the signal hemisphere must 
be identified as pion candidates.
The \piz candidates are reconstructed from 
two separate EMC clusters with energies above 100 MeV that are 
not associated with charged tracks; these \piz candidates are 
required to have
invariant masses 
between 100 and  160 \MeVcc.
Events are required to have a single \piz in the signal hemisphere.
The \ta candidates are reconstructed in the signal hemisphere
using the three tracks and the \piz candidate. The invariant 
mass of the \ta candidate, $\mtpz$, is required to be less than 
the mass of the \ta lepton.
The Monte Carlo simulation predicts that 14\% of the events 
remaining after the event selection process are \ta-pair 
events that do not contain a \tppppz decay, and 1.3\% are
\epem\to\qqbar events.

\begin{figure}[tb]
\begin{center}
  \includegraphics[width=0.45\textwidth]{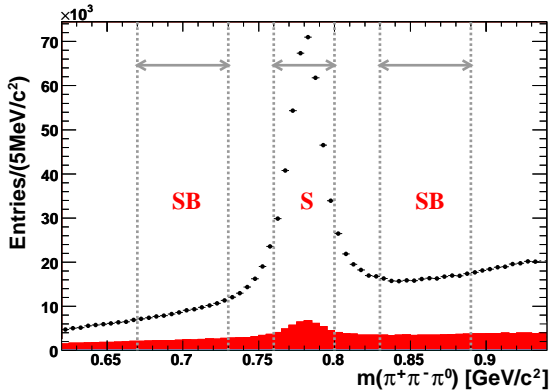}
\end{center}
\caption{\om candidate mass spectra for selected events in data and 
  background expected from simulation (shaded histogram).  The background
  histogram does not include the non-resonant \tppppz decays.
  The signal (S) and sideband (SB) regions are indicated in the figure.}
\label{fig:ommass}
\end{figure}

Each selected event has two \pppz\ combinations.
The \om signal
region is defined for masses $m(\pppz)$ between 760 \MeVcc and 800 \MeVcc with
mass regions of width 60 \MeVcc on each side of the peak used
as sideband regions for background studies, as shown in 
Figure~\ref{fig:ommass}.
For each \om\ candidate, the angle $\thet$ 
is calculated. The distribution of $\cos \thet$ is used for
the SCC measurement.

There are three background sources to be considered in this analysis:
combinatoric background, \qqbar events and non-signal $\tau$ decays.
The combinatoric background
is expected, and confirmed by the simulation, to have a  
distribution of $\cos \thet$ which is 
independent of \mopz. This allows the sideband regions to be used
to subtract this background. 
The number of combinatoric events lying within the signal region 
is obtained by fitting the \mopz spectrum with a relativistic
Breit-Wigner convolved with a resolution function 
for the \om resonance and a polynomial for the 
combinatoric background.  The polynomial is integrated over the 
signal region to find the number of continuum events in the 
signal region. 

After subtracting the combinatoric background, 
approximately 0.3\% of the remaining
events in the signal region are expected to be of \qqbar origin, 
while 4.6\% are expected to be from non-signal \ta decays.
The background from \epem\to\qqbar\ events that contain \ompppz decays
is studied using events with $m(\ppppz)$ 
well above the \ta mass ($>2.1\gevcc$).
In this region, where all events are considered to be of \qqbar origin,
a comparison of the numbers of \om mesons in simulation and data is used 
to 
scale the simulated \qqbar background before subtracting from data.

The dominant non-signal \ta background, comprising 99\% of the
remaining background, is \tauomppz, where the additional \piz has 
not been reconstructed.
The decay \tauomppz has not been well measured and is 
incorrectly modeled in the Monte Carlo. 
To correct for the differences between data and Monte Carlo,
events with an additional \piz candidate in the signal hemisphere
are selected, using the same cuts discussed above.
Using these events, the Monte Carlo branching fraction of \tauomppz is 
corrected by comparing the numbers of fitted \om candidates in 
data and Monte Carlo. The fit function used for this is a 
relativistic Breit-Wigner convolved with a resolution function 
plus a polynomial background.
The branching fraction obtained using this correction
technique is found to be consistent 
with existing measurements \cite{pdg}.
To correct the angular distribution of \tauomppz,
backgrounds, consisting of combinatorics, \qqbar events and 
\tauom decays (assuming the angular distribution 
corresponding to the dominant FCC contribution),
are subtracted from the \omppz data sample, and
the remaining \costhet distribution is used to correct
the \tauomppz distribution in the Monte Carlo. 

After subtracting backgrounds and applying \costhet-dependent 
efficiency corrections, 
a binned fit to the remaining \costhet\ distribution is carried out using 
\begin{equation}
  F(\costhet)\,
  =N\times\left[\frac{1}{2} \epsilon + 
\frac{3}{4}\left(1-\epsilon \right)\left(1-\mbox{cos}^2\thet\right)\right],
\label{eq:onescc}
\end{equation}
where $N$ is a normalization factor and 
the parameter $\epsilon$ is the fraction 
of \tauom decays that proceed through SCC.
In Eq.\ref{eq:onescc}, only the $L=0$ term is used to describe
the SCC contribution since this
function gives the most conservative estimate of \epsil
(i.e. the largest upper limit).

To help validate the analysis method, the procedures were 
applied to Monte Carlo samples generated to include small
fractions (1\% and 2\%) of the second-class current 
process $\tau^- \to b_1(1235)^-\nu_\tau \to \omega\pi^-\nu_\tau$,
as well as to a sample containing no SCC contribution. In each
case the fits to the simulated detector-level, 
background-subtracted angular  distributions returned values 
consistent with the input fractions of SCC.

The largest contributions to systematic uncertainties on \epsil are
scaling and modeling the Monte Carlo background. The correction applied 
to the branching fraction of \tauomppz has an error associated with 
it, determined by the available statistics. This correction factor 
is adjusted by $\pm1\sigma$ to obtain
the uncertainty in \epsil while the errors associated with 
correcting the angular distribution are folded into the statistical
uncertainty. In addition, there are \ta decays that may be present in 
the final event sample but which are not included in the simulation. The 
largest 
of these are expected to be $\tm\to\omega K^-\nut$,
$\tm\to\omega\pim\piz\piz\nut$, and $\tm\to\omega \pim\pim\pip\nut$ 
decays 
which, when combined, can add up to 0.2\% of the final event sample.
Since the effect that these decays have on the angular distribution
is unknown, the extreme cases are taken to obtain the uncertainty.
These cases correspond to the decays having either entirely 
$1-\cos^2\thet$ or entirely $\cos^2\thet$ distributions.
The scaling of \qqbar events can also affect the measurement 
of \epsil, and the uncertainty is obtained by adjusting the scaling
factor by $\pm1\sigma$.
These systematic uncertainties are summarized in Table \ref{table:syst}.

\begin{table}[ht]
\caption{Summary of systematic uncertainties on \epsil}
\begin{center}
\begin{tabular}{c|c}
\hline
Source & Uncertainty ($\sigma_\epsil$) \\
\hline
\Br(\tauomppz) & $\pm0.07\%$\\
non-simulated \ta decays & $_{-0.55}^{+0.00}\%$ \\
\qqbar scaling & $\pm0.01\%$\\
\hline
Total & $_{-0.55}^{+0.08}\%$ \\
\hline
\end{tabular}
\end{center}
\label{table:syst}
\end{table}

Subsets of the generic Monte Carlo dataset are used in the background 
studies and in the determination of the efficiencies. Therefore,
to estimate the sensitivity of the analysis without the effect of 
statistical correlations in the Monte Carlo samples, 
an ensemble of simulated experiments is used. 
In this study, angular distributions are
generated for the signal and sideband regions to simulate the statistics 
available in the data and various Monte Carlo samples used in the 
analysis, with $\epsilon = 0$ in the signal Monte Carlo.  
After subtracting background samples, the 
angular distribution is corrected for efficiency and fitted
using Eq.~\ref{eq:onescc}.
The statistical uncertainty on \epsil obtained from the fit is 
$0.63\%$, which combined with the systematic uncertainties
leads to an estimated uncertainty of 
$\sigma_\epsil = _{-0.84}^{+0.64}\%$.

\begin{figure}[hbt]
\begin{center}
  \includegraphics[width=0.45\textwidth]{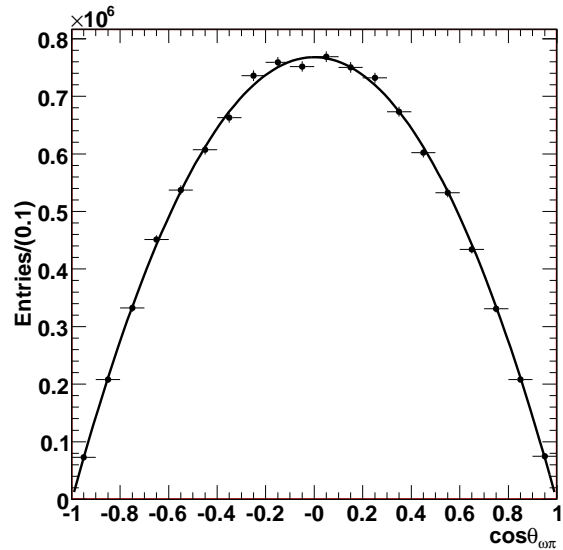}
\end{center}
\caption{
  The \costhet distribution for the data. The curve is
  the result of the fit described in the text.}
\label{fig:dataCosth}
\end{figure}

The angular distribution of the final state particles
in data is obtained by subtracting estimated backgrounds
as described above.
The remaining distribution is corrected for efficiency and fitted using
Eq.~\ref{eq:onescc} as shown in Figure \ref{fig:dataCosth}. 
The fit has $\chi^2/dof =$ 15.4/18, and the fitted value of
\epsil in the data is \datEps, which is consistent with no SCC 
contribution to \tauom decays.

The upper limit on $\epsilon$ is obtained using 
a Bayesian approach~\cite{bayesian} with a prior that is 
flat for $\epsil > 0$ and zero for $\epsil < 0$.
The probability distribution for the value of the SCC contribution
is a Gaussian with mean $\epsilon=-0.55\%$ 
and errors $\sigma_\epsilon=_{-0.84}^{+0.64}\%$, taken from
the simulation studies; however
since negative values of \epsil are non-physical, 
only the positive 
portion of this probability distribution is used in the limit calculation. 
The limits obtained from this method 
are $\epsilon <$~\thislimitEpsBayes at 90\% CL 
and $\epsilon <$~\thislimitMoreEpsBayes at 95\% CL.

In summary, a search for second-class currents in the decay \tauom 
has been 
conducted with the \babar\ detector. 
No evidence for second-class currents is observed, 
and a 90\% confidence level Bayesian upper limit for 
the fraction of the second-class current in \tauom decays is
set at \thislimitEpsBayes. 
For comparison with the previous result from CLEO,
this is equivalent to a ratio of second-class (non-vector)
to first-class (vector) currents of \thislimitRatBayes.
This limit is an order of magnitude lower than
the limit set by the CLEO collaboration \cite{cleoscc}.

\section{Acknowledgments}
\label{sec:acknowledgments}
We are grateful for the excellent luminosity and machine conditions
provided by our \pep2\ colleagues, 
and for the substantial dedicated effort from
the computing organizations that support \babar.
The collaborating institutions wish to thank 
SLAC for its support and kind hospitality. 
This work is supported by
DOE
and NSF (USA),
NSERC (Canada),
CEA and
CNRS-IN2P3
(France),
BMBF and DFG
(Germany),
INFN (Italy),
FOM (The Netherlands),
NFR (Norway),
MES (Russia),
MEC (Spain), and
STFC (United Kingdom). 
Individuals have received support from the
Marie Curie EIF (European Union) and
the A.~P.~Sloan Foundation.


\begin{thebibliography}{99}
\bibitem{charge}
Charge-conjugate reactions are implied throughout this paper.
\end{thebibliography}
\end{document}